\documentclass[preprint,epsfig,aps,showpacs]{revtex4}
\usepackage{amsmath}
\usepackage{graphicx}

\newcommand{\f}{\begin{equation}}
\newcommand{\ff}{\end{equation}}
\newcommand{\fa}{\begin{eqnarray}}
\newcommand{\ffa}{\end{eqnarray}}

\begin{document}
\title{The power-law expansion universe and the late-time behavior}
\author{Yi-Huan Wei${ }^{1,2}$}
\affiliation{%
${ }^1$ Department of Physics, Bohai University, Jinzhou 121000,
 Liaoning, China}
\affiliation{%
$ { }^2$ Institute of Theoretical Physics, Chinese Academy of
Science, Beijing 100080, China}
\begin{abstract}
~~ Using the SNe Ia data we determine the three parameters in the
power-law expanding universe model with time-dependent power
\cite{W}. Inputting $H_0$ and $t_0$, then we find the $\dot{a}-t$
evolution curve with $m=5.0$ and $q_0=-0.90$ can fit very well to
that from SNe observation data. The model predicts the transition
redshift $z\simeq0.38$. The dark energy deduced from this model
have phantom property but the universe doesn't encounter the Big
Rip singularity. Assuming that this model with the three
parameters is valid for the future universe, then we predict that
the total energy density of the universe is decreasing and will
soon reach its minimum.

\end{abstract}

\maketitle

\section{Introduction}
The supernovae Type Ia (SNe Ia) observation and the cosmic
microwave background power spectrum measurement show the existence
of dark energy and the flatness of the universe \cite{Riess,Ber}.
In order to understand the nature of dark energy, many dynamical
models have been proposed, such as, quintessence, tachyon,
k-essence, etc.
\cite{CRS,K,T,C,Caldwell,P,A,Sahni,Peebles,Wei,Li}. A cosmological
constant is the simplest model of dark energy, but no natural
explanation for the origin of it can be given. The SNe Ia
observations provide the currently most direct way to probe the
dark energy through the luminosity-distance relation
\cite{Rie,Spergel,Tonry,Daly,Chou,Alam,Knop,Zhu,Yun,Abramo,Dicus,Chae,Dev,Bertolami}.

In order to probe the evolution of dark energy, an appropriate and
simple parametrization of them can take the advantage of being
practical \cite{Jassal}. Despite the broad interest in proposing
dynamical model of dark energy, their physical properties are
still poorly understood at a fundamental level. As well known,
almost all models of dark energy can fit to the observation data,
but few models fitting well to the evolution of the universe in a
very long history have been found. From a phenomenological point
of view all known models of dark energy should be equivalent. We
still have a long way from being able to give a complete
explanation for the nature of dark energy. Nevertheless, It is
possible to try to seek an appropriate form of solution to
Friedmann equation independent of the concrete dark energy model.

Recently, we propose a power-law expansion universe model
\cite{W}, $a=a_0t^{n(t)}$ with $n(t)=n_0+bt^m$, which for a
spatially flat, isotropic and homogeneous universe leads to the
consistent results with those given in some current researches
such as \cite{Yun}. Here, the parameters $m$, $n_0$ and $b$ will
be determined by fitting the evolution of the universe from the
SNe data. The three cases are discussed, among which the best one
has the parameters $m=5$, $q_0=-0.9$ and
$b=2.76671\times10^{-54}$. The model with the three best
parameters predicts $z_T\simeq0.38$, favors the phantom dark
energy for the current universe, and show the total energy density
of the universe is decreasing.

\section{Determination of parameters by SNe data}
In order to describe the the universe transition from deceleration
to acceleration expansion we propose such a universe model
described by
\begin{eqnarray}
a=a_0t^{n}=a_0t^{n_0+bt^m}, \label{scale}
\end{eqnarray}
where $a_0$ is the scale factor for $t=1yr$, $n_0$, $b$ and $m$
are three nonnegative parameters \cite{W}. From (\ref{scale}),
there is
\begin{eqnarray}
\dot{a}=a(\dot{n}\ln t+\frac{n}{t})=\frac{a}{t}[n_0+bt^m(1+mlnt)],
\label{eq2}
\end{eqnarray}
where a dot stands for the derivative with respect to $t$. Giving
the observed quantities, the current Hubble parameter $H_0$, the
current deceleration parameter $q_0$ and the age of the universe
$t_0$, then we have
\begin{eqnarray}
n_0=H_0t_0-b(mx_0+1)t_0^m, \quad
b=\frac{1-(q_0+1)H_0t_0}{m(mx_0+2)}H_0t_0^{1-m}. \label{eq3}
\end{eqnarray}
In order to determine $n_0$ and $b$, one needs to know
$H_0=H(t_0)$, and choose the parameters $q_0=q(t_0)$ and $m$.
Fixing $H_0$ and $t_0$, using the different values of $q_0$ and
$m$, then one can obtain the diffeent $n_0$ and $b$, which yield
different transition redshift $z_T$. By adopting $z_T=0.5$, then
we can determine the values of $m$, $n_0$ and $b$ \cite{W}.

\begin {figure}
\begin{center}
\includegraphics[width=3.9in, height=3.3in]{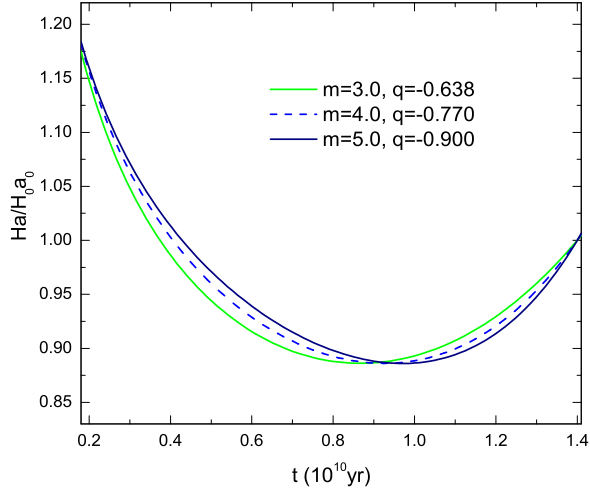}
\caption{The $\dot{a}/a_0H_0-t$ figure for three special cases
with different values of $m$ and $q_0$.} \label{F1}
\end{center}
\end{figure}
\begin{figure}
\begin{center}
\begin{tabular}{ccc}
\includegraphics[width=2in, height=2.3in]{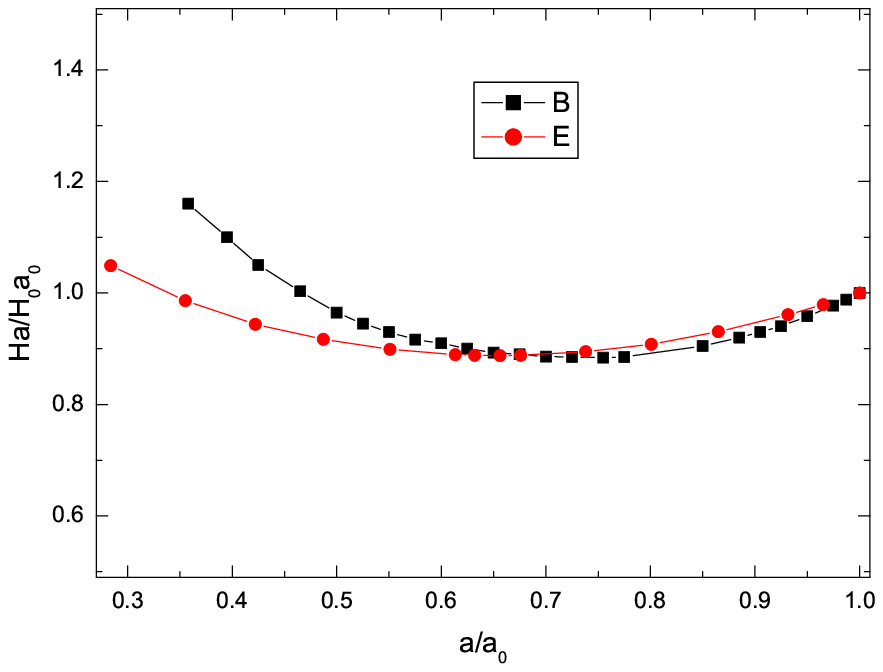}& \includegraphics[width=2in,
height=2.3in]{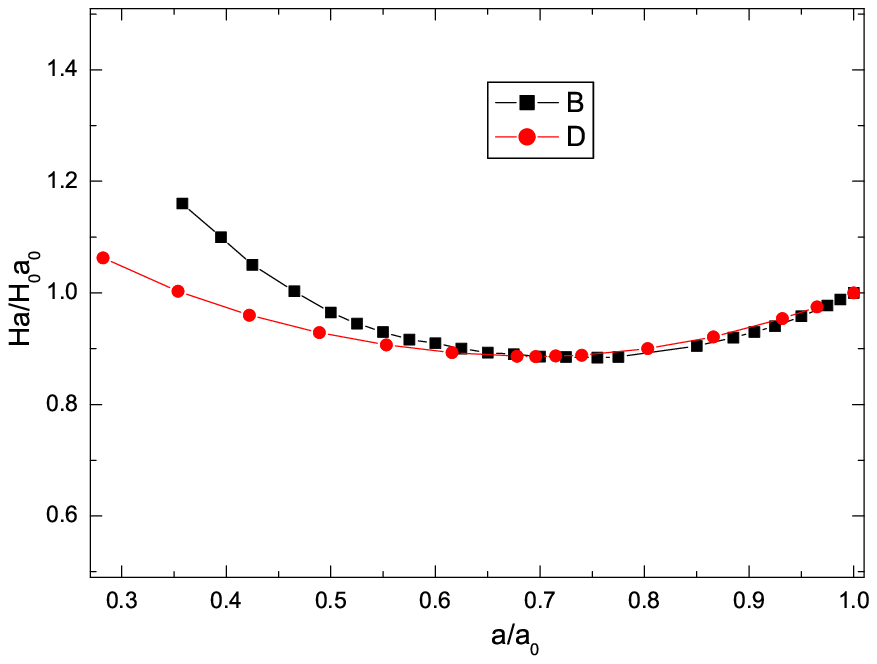}&
\includegraphics[width=2in, height=2.3in]{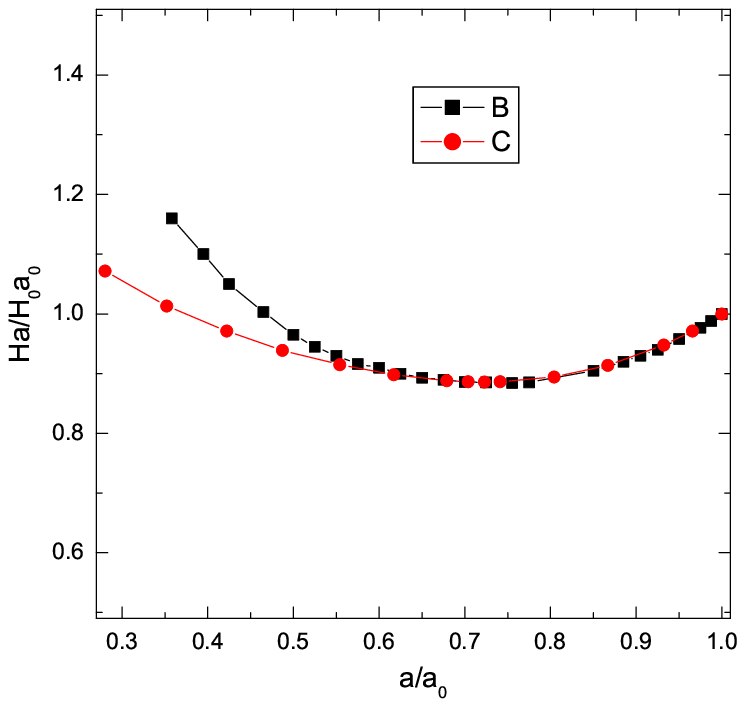}
\\
\qquad(a)  & \qquad(b)  & \qquad(c)
\\
\end{tabular}
\caption{The $\dot{a}/a_0H_0-a/a_0$ figure. Curve B comes from the
supernovae data set consisting of the gold sample (Ref.
\cite{Chou}). E, D and C in (a), (b) and (c) correspond to the
three curves in FIG. \ref{F1}, ($q_0=-0.638$, $m=3.0$),
($q_0=-0.77$,$m=4.0$) and ($q_0=-0.90$,$m=5.0$), respectively,}
\label{F2}
\end{center}
\end{figure}

Here, we will still fix $H_0$ and $t_0$, and treat $m$ and
$q_0=q(t_0)$ as two input parameters. The improvement to our
previous method is that we will not only require the model can
describe the universe transition but hope that it may track the
histoty of the universe evolution. From equation (\ref{eq2}), one
can have
\begin{eqnarray}
\dot{a}/a_0H_0=\frac{a}{a_0H_0t}[n_0+bt^m(1+mlnt)].
\label{eq-adot}
\end{eqnarray}

FIG. \ref{F1} shows $\dot{a}/H_0-t$ curves with the same minimum
for the three special cases with different values of $m$ and
$q_0$. FIG. \ref{F2} gives the comparison between $\dot{a}/H_0-a$
curves given from the SNe Ia data \cite{Chou,Rie} and  FIG.
\ref{F1}. We find that among the three cases the third one (c),
i.e., the curve with $m=5.0$ and $q_0=-0.9$ can best fit to the
evolution curve from the SNe Ia data. This implies that the model
predicts the deceleration parameter of the universe $q_0=-0.90$.
This value is bigger than some known results, such as, $q_0 = 0.35
\pm0.15$ given in \cite{Da}, but may be consistent with the
results in \cite{Cardone,Gong}.

Now, we determine the universe transition redshift. Putting
$m=5.0$ and $q_0=-0.90$ in equation (\ref{eq3}) yields
$b=2.76671\times10^{-54}$, $n_0=0.804696$, $A=2n_0+2m-1=10.6094$
and $B=m(2n_0+m-1)=28.047$. Using equation (8) in \cite{W} and
letting $q=0$, then we obtain
\begin{eqnarray}
b^2(1+mx_T)^2t_T^{2m-2}+b(A+ Bx_T)t_T^{m-2}+(n_0^2-n_0)t_T^{-2}=0,
\label{eq-q}
\end{eqnarray}
where $x_T=lnt_T$ and $t_T$ denotes the transition time. Equation
(\ref{eq-q}) yields $t_T\simeq 9.71Gyr$, and from the relation
$z=\frac{a(0)}{a(t)}-1$ we have the transition redshift
\begin{eqnarray}
z_T=\frac{a(t_0)}{a(t_T)}-1=\frac{t_0^{n_0+bt_0^m}}{t_T^{n_0+bt_T^m}}-1\simeq0.383,
\label{eq-z}
\end{eqnarray}
which may be consistent with some known results, such as those
given in \cite{Alam,Czaja,Rie}.

In this section, by fitting to the SNe data we determine the
parameters in the power-law expanding universe model with ($m=5$,
$n_0=0.804696$, $b=2.76671\times10^{-54}$) . Next, using the scale
factor (\ref{scale}) with known parameters $m$, $n_0$ and $b$ we
will catch a glimpse of the evolution of dark energy and the
universe.

\section{Evolution properties of the universe}
For the spatially flat, isotropic and homogeneous universe
described by the scale factor (\ref{scale}), the total energy
density is determined by
\begin{eqnarray}
\rho=\frac{3}{M_{P}^2}H^2, \quad H=\frac{\dot{a}}{a},
\label{eq-Hubble}
\end{eqnarray}
with $M_P=1/\sqrt{8\pi G}=\frac{3H_0^2}{\rho_0}$ the reduced
Planck mass. Assuming that the matter component is the perfect
fluid, i.e., $\rho_m=\rho_0a(t_0)^3/a(t)^3$, then the dark energy
density is given by
\begin{eqnarray}
\rho_X=\rho_0[(\frac{H}{H_0})^2-a(t_0)^3/a(t)^3]. \label{eq-dark}
\end{eqnarray}
\begin {figure}
\begin{center}
\includegraphics[width=3.9in, height=3.5in]{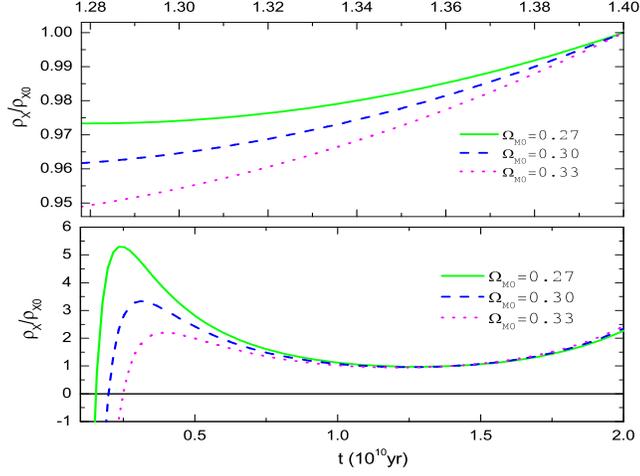}
\caption{The $\rho_X-t$ figure is given for $m=0.5$ and $q=-0.9$,
which shows the evolution for dark energy in the past epoch for
$\Omega_{m0}=0.27,0.30,0.33$, respectively.} \label{F3}
\end{center}
\end{figure}
\begin {figure}
\begin{center}
\includegraphics{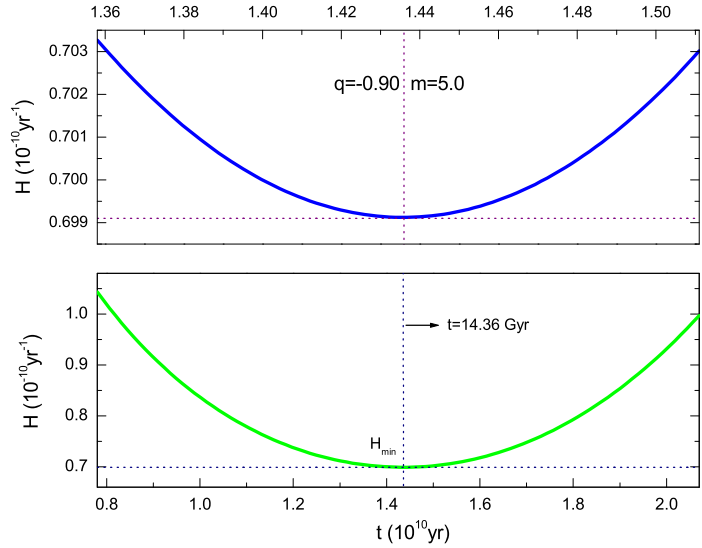}
\caption{The $H-t$ figure shows the Hubble parameter has the
minimum $H_{min}\simeq 0.6991\times10^{-10}yr^{-1}$ at $t\simeq
14.36$ $Gyr$.} \label{F4}
\end{center}
\end{figure}
\begin {figure}
\begin{center}
\includegraphics{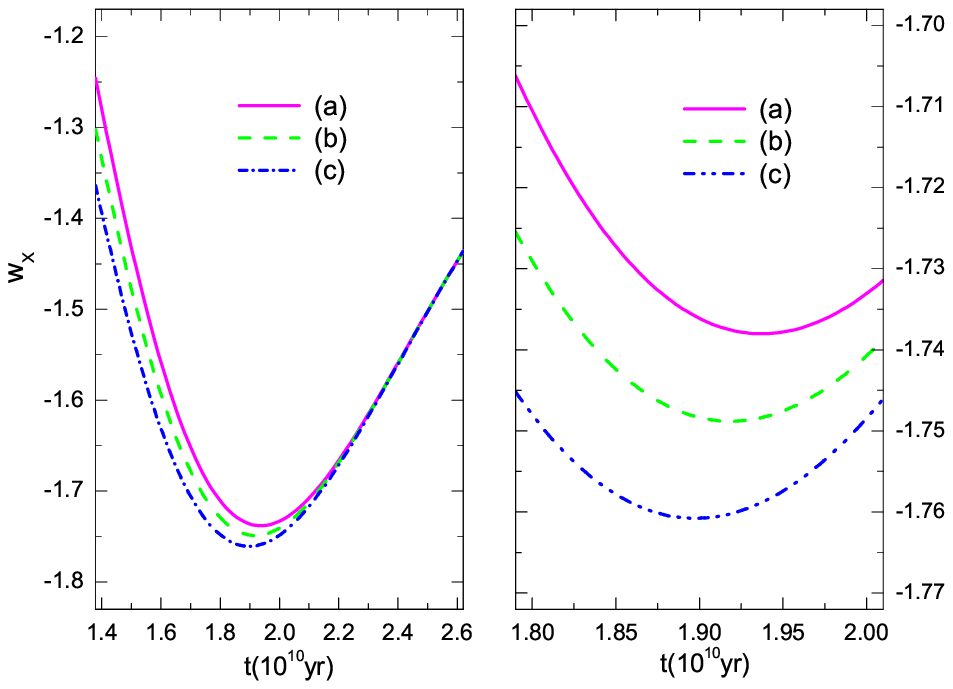}
\caption{Curves $(a)$,$(b)$ and $(c)$ denote $w_X$ for
$\Omega_{m0}=0.27,0.30,0.33$, respectively. The figure shows $w_X$
will reach its minimum value at $t=18.5\sim 19.5$ $Gyr$.}
\label{F5}
\end{center}
\end{figure}
FIG. 3 shows the phantom property of dark energy at the current
epoch and FIG. 4 shows the minimum of Hubble parameter
$H_{min}\simeq0.699\times10^{-10}$ $yr^{-1}$ at $t\simeq14.36$
$Gyr$, which yields the minimum of universe energy density
$\rho_{min}=(\frac{H_{min}}{H_0})^2\rho_0\simeq 0.997\rho_0$.

From the conserved equation for dark energy
\begin{eqnarray}
\dot{\rho}_X+3H(\rho_X+p_X)=0, \label{con}
\end{eqnarray}
where a dot denotes the derivative with respect to time, one can
obtain the equation of state
\begin{eqnarray}
w_X=\frac{p_X}{\rho_X}=-\frac{\dot{\rho}_X}{3H\rho_X}-1.
\label{wx}
\end{eqnarray}
Assuming that the dark energy is slowly changing, then around the
current epoch we can approximately have
$w_X\simeq-\frac{\bigtriangleup{\rho}_X}{3H_0\rho_{X0}\bigtriangleup
t}-1$, where  $\bigtriangleup{\rho}_X=\rho_{X0}-\rho_X$ and
$\bigtriangleup t=t_0-t$. For $t=1.399\times10^{10}$ $yr$, there
are the energy density ratio
$\rho_X/\rho_{X0}=0.99956,0.99946,0.99936$, which lead to
$w_{X0}=-1.21,-1.26,-1.30$ for $\Omega_{m0}=0.27,0.30,0.33$,
respectively.

From the Friedmann equations
\begin{eqnarray}
H^2=\frac{1}{3M_P^{2}}(\rho_X+\rho_m), \quad
H^2+\dot{H}=-\frac{1}{6M_P^{2}}(\rho_X+\rho_m+3p_X), \label{eq11}
\end{eqnarray}
one can have
\begin{eqnarray}
w_X=\frac{p_X}{\rho_X}=-\frac{3H^2+2\dot{H}}{3H^2-\rho_m/M_P^2},
\label{eqde}
\end{eqnarray}
with $\dot{H}=b[m(m-1)\ln{t}+2m-1]t^{m-2}-n_0t^{-2}$ and
$H=bt^{m-1}(1+mlnt)+n_0t^{-1}$. FIG. 5 illustrates the future
evolution of dark energy, which will evolve to its minimum in the
time interval $18.5\sim19.5$ $Gyr$, such as,
$w_{Xmin}\simeq-1.738$ at $t\simeq19.36$ $Gyr$ for
$\Omega_{m0}=0.27$. For $t\gg1$, $\rho_m$ may be neglected and for
$m=5$ equation (\ref{eqde}) reduces to
\begin{eqnarray}
w_X\simeq-1-\frac{8}{15b}(t^5lnt)^{-1}, \label{eq13}
\end{eqnarray}
which implies that $w_X$ decreases rapidly to $-1$ for late time.

For the constant equation of state $w_X<-1$, the universe will
encounter the sudden future singularity \cite{Caldwell}, i.e., Big
Rip or Big Smash, but it may evade a Big Rip if the the equation
of state falls off quickly \cite{Wei}. Clearly, the power-law
expansion universe model considered currently describe such a
phantom universe without the future singularity.

We study the power-law model with time-dependent power proposed in
\cite{W} by using the SNe data. For $m=5$, $q_0=-0.9$ and
$b=2.76671\times10^{-54}$, the model can well track the evolution
of the supernova to a very high redshift, and the predicted
transition reashift $z_T$ and deceleration parameter $q_0$ may be
consistent with some known results
\cite{Rie,Da,Cardone,Gong,Czaja}. The model may be expected to be
able to describe the future evolution of the universe in the near
future since it can well track the past evolution history of the
universe. Considering the matter as the perfect fluid, then the
model predicts that the equation of state parameter for dark
energy is decreasing and will go to its minimum value after about
$5$ $Gyr$ for $\Omega_{m0}\sim0.3$. Another interesting result
given by the model is that the total energy density of universe is
dropping and will soon approach its minimum value smaller slightly
than the current one.

Obtaining a overall scale factor from a dark energy model should
be what everyone yearns, who is working in this research realm.
However, it is a pity that there are few models which can track a
very long evolution history of the universe by fitting the
observation data at one point. So, studying the scale factor in a
direct way doesn't interpret the nature of dark energy but should
be very helpful for understanding the behavior of dark energy.

 \vskip 1.9cm

{\bf Acknowledgement:} This work are supported by ITP Post-Doctor
Project 22B580, Chinese Academy of Science of China, National
Nature Science Foundation of China and Liaoning Province
Educational Committee Research Project.

\end{document}